\documentclass{article}

\usepackage{graphicx}      
\usepackage{amsmath,amssymb,mathrsfs}
\usepackage{enumerate}
\usepackage{bm}
\newtheorem{theorem}{Theorem}
\newtheorem{remark}{Remark}
\newtheorem{lemma}{Lemma}
\newtheorem{corol}{Corollary}
\newtheorem{definition}{Definition}

\begin{document}
\title{Tracking performance of PID for nonlinear stochastic systems}

\author{Cheng~Zhao and Shuo~Yuan\footnote{Cheng Zhao and Shuo Yuan are with the Key Laboratory of Systems and Control,
Academy of Mathematics and Systems Science, Chinese Academy of
Sciences, Beijing 100190, China(e-mail: zhaocheng@amss.ac.cn; syuan@amss.ac.cn). Corresponding author: Shuo Yuan.}}
\date{}
\maketitle

\begin{abstract}In this paper, we will consider a class of continuous-time stochastic control systems with both unknown nonlinear structure and unknown disturbances, and investigate the capability of the classical proportional-integral-derivative(PID) controller in tracking time-varying reference signals. First, under some suitable conditions on system nonlinear functions, reference signals, and unknown disturbances, we will show that PID controllers can be designed to globally stabilize such systems and ensure the boundedness of the tracking error. Analytic design formulae for PID gain matrices are also provided, which only involve some prior knowledge of the partial derivatives of system structural nonlinear functions. Besides,  it will be shown that the steady-state tracking error hinges on three critical factors: i) the change rate of reference signals and external disturbances; ii) the intensity of random noises; iii) the selection of PID gains, and can be made arbitrarily small by choosing PID gains suitably large. Finally, by introducing a desired transient process which is shaped from the reference signal, we will present a new PID tuning rule, which can guarantee both nice steady-state and superior transient control performances.
\newline

\end{abstract}


\section{Introduction}Feedback is a basic concept in automatic control, whose primary objective is to reduce the effects of the plant uncertainty
on the desired control performance. The PID controller is by far the most basic and  dominating form of feedback in use today, and is believed to be
``bread and butter" of control engineering \cite{As}. In fact, more than 90\% of all control loops are PID \cite{As}. Of course, one of the main
challenging tasks for the implementation of the PID controller is how to design the three controller parameters, which has been investigated extensively
 in the literature, e.g., for linear time-invariant systems with time delay \cite{Ma2019,Appeltans2022},  for affine nonlinear systems with adaptively
 tuning PID gains \cite{song2017}, and for mechanical systems with special structures \cite{Romero,Takegaki}, just to name a few.

As the most widely adopted control method in industrial processes, the PID controller is favored by various control practitioners because of its simple
structure and ease of implementation. However, there are limited theoretical investigations explaining the rationale for why PID is so successful in
dealing with various practical nonlinear systems with uncertainties. In recent years, some rigorous mathematical studies on the theory and design of
classical PID have been made for uncertain nonlinear systems, see e.g., \cite{Zhao2017,2022,Zhang2022,lguo}, which have shown explicitly that PID
control has the ability to deal with a vast range of system structural uncertainties. For instance, in \cite{zhao2016}, the authors have shown that
the classical PID control has the ability to globally stabilize a class of nonlinear uncertain systems without control channel uncertainty, and
make the system output converge to any given constant setpoint, provided that some prior knowledge of the Lipschitz constants of system nonlinear
functions are available.  Moreover, necessary and sufficient conditions for the selection of the PID parameters have also been provided for a class
of single-input-single-output(SISO) affine nonlinear systems  in \cite{Zhao2017}. To the best of our knowledge, the global stability of a class of
PID controlled nonaffine MIMO uncertain nonlinear systems was established for the first time in \cite{2022}. Then, the authors further extended the
deterministic results to stochastic cases in \cite{Zhang2022}.

It is worth mentioning that most of the investigations on the analysis and design of PID control assume that the reference signal is
time-invariant(or converges to a constant), and do not account for unknown external uncertainties, see e.g.
\cite{Khalil,Chang,Zhao2017,zhao2016,2022,Zhang2022}. For theoretical studies of PID controlled nonlinear stochastic systems, the diffusion term is
 usually required to vanish at the setpoint(see e.g., \cite{cong2021,Zhang2022}), which is also an unrealistic assumption for many practical systems.
  In order to further reveal the rationale of the linear PID control, it is necessary to take both internal structural uncertainties and unknown
  external disturbances into consideration. Also, it is of vital importance to investigate the capability of PID in dealing with general time-varying
   reference signals. These facts motivate us to consider a class of nonlinear stochastic systems subject to both unknown dynamics and external
   disturbances, where both the reference signal and the external disturbance are not assumed to be slowly time-varying. Besides, the diffusion
   term is not required to vanish at the reference signal in this paper.

The contributions are summarized as follows:
\noindent\begin{itemize}
\item First, we will show that the classical PID control can globally stabilize such systems and ensure the boundedness of the tracking error under
suitable conditions on system nonlinear functions, disturbances, and reference signals. Also, analytic design formulae for PID gain matrices are
provided, which are only based on bounds of partial derivatives of the nonlinear drift and diffusion functions.
\item It will be shown that the steady-state error is determined by the change rate of reference signals and external disturbances,  the intensity of
the stochastic disturbances, and choices of the PID gain matrices. Moreover, it will be shown that the PID gains can be chosen large enough to achieve
practical tracking, even if the reference signal is not slowly varying.
\item  A new PID tuning rule is also presented, which can guarantee both nice steady-state and superior transient control performance of the closed-loop
 systems.
\end{itemize}

The remainder of this paper is organized as follows. In
Section 2, we introduce the problem formulation, including the notations, the control systems, and the assumptions. Section 3 presents the main results
 of this paper, with their proofs given in Section 4.   Finally, we conclude the paper with some remarks in Section 5.

\section{\label{sec:P}Problem Formulation}
\subsection{Notations}
Denote $\!\mathbb{R}^{m\times n}\!$ as the space of $m\!\times\! n$ real matrices,  $|x|$  as the Euclidean norm of a vector $x$, $A^{\top}$ as the transpose of a matrix $A$. Let  $\mathbb{R}^+=[0,\infty)$. The norm of a matrix $P \in \mathbb{R}^{m\times n}$  is defined by $\|P\|=\sup_{x\in\mathbb{R}^n, |x|=1}|Px|$. For a symmetric matrix $S$, let $\lambda_{\mathrm{min}}(S)$ and $\lambda_{\mathrm{max}}(S)$  denote the smallest and largest eigenvalues of $S$, respectively.
 For two symmetric matrices $S_1$ and $S_2$ in $\mathbb{R}^{n\times n}$, the notation $S_1\geq S_2$(or $S_2\le S_1$) means  that $S_1-S_2$ is a positive semi-definite matrix.
For a function $f(x_1,x_2\cdots,x_k)\in C^{1}(\mathbb{R}^{n_1}\times\mathbb{R}^{n_2}\times\cdots\times\mathbb{R}^{n_k},\mathbb{R}^{m})$, let
$\frac{\partial f}{\partial {x_i}}(x_1,\cdots,x_k)$ denote  the $m\times n_i$ Jacobian matrix of $f$ with respect to $x_i$ at the point $(x_1,\cdots,x_k)$. For a random variable $X$, let $\mathbf E(X)$ denote its expectation.

\subsection{The control systems}
Consider a class of nonlinear stochastic systems with both unknown nonlinear structure and unknown disturbance:
\begin{align}\label{sys}\begin{split}\begin{cases}
		\mathrm{d}x_1(t)=&\!\!x_2(t) \mathrm{d}t\\
		\mathrm{d}x_2(t)=&\!\!\left[f(x_1(t),x_2(t),u(t)) +d(t)\right]\mathrm{d}t +g(x_1(t),x_2(t)) \mathrm{d}B_t,\end{cases}
\end{split}
\end{align}
where $(x_1,x_2)\in\mathbb{R}^{2n}$ is the state vector, $u\in\mathbb{R}^{n}$ is the control input, $f\in C^1(\mathbb{R}^{3n}, \mathbb{R}^{n})$ and $g\in C^1(\mathbb{R}^{2n}, \mathbb{R}^{n})$ are system structural nonlinear functions, $d\in C^1(\mathbb{R}^+,\mathbb{R}^n)$ is the unknown external disturbance and $\{B_t,t\geq 0\}$ is a one-dimensional standard Brownian motion.
\begin{remark}
Many physical systems can be described by model (\ref{sys}) via the Newton's second law in mechanics or some fundamental physical laws in electromagnetics. For instance, it can  be used to describe the motion of a controlled moving body in $\mathbb{R}^n$. The well-known Langevin equation can also be described by model (\ref{sys}). Besides, note that unknown dynamics, external disturbances, and random noises are simultaneously taken into account in system (\ref{sys}),  which makes it applicable to many practical processes including motion control systems and power systems, etc.
\end{remark}

The control objective is to globally stabilize the uncertain stochastic system  (\ref{sys}) in the mean square sense and  to make the controlled variable $x_1(t)$ track a given \emph{time-varying} reference signal $r(t)$.

The classical PID controller is given by
\begin{align}\label{control}\begin{split}
	u(t)=&K_{p} e(t)+K_{i} \!\int_{0}^{t}\! e(s)\mathrm{d}s+K_{d} \dot{e}(t),\\
e(t)=&r(t)-x_1(t),\end{split}
\end{align}
where $e(t)$ is the tracking error, $K_p$, $K_i$ and $K_d$ are $n\times n$ PID gain matrices to be designed. In this paper, we aim to investigate the capability of the classical PID (\ref{control}) in dealing with the nonlinear stochastic system  (\ref{sys}), and focus on the tracking performance of the closed-loop system.
\subsection{Assumptions}
We first introduce some assumptions about the uncertain nonlinear functions, the time-varying reference signal and the external disturbance.

\noindent \textbf{Assumption 1.} The nonlinear functions $f$ and $g$ satisfy
\begin{align*}
&\Big\|\frac{\partial f}{\partial x_i}\Big\| \le L_i,~
\Big\|\frac{\partial g}{\partial x_i}\Big\| \le N_i,~i=1,2,~\text{for all~}x_1,x_2,u\in\mathbb{R}^n,
\end{align*}
where $L_1,L_2, N_1$ and $N_2$  are positive constants. Besides,
\begin{align*}
0<\underline bI_n\le  \frac{\partial f}{\partial u}\le \bar b I_n,~\text{for all~}x_1,x_2,u\in\mathbb{R}^n,
\end{align*}
where $0<\underline b\le \bar b$ are constants and $I_n$ is the identity matrix.

\begin{remark}We remark that the boundedness of the partial derivatives $\frac{\partial f}{\partial x_i}$($i=1,2$) appears to be necessary in general for global results, since the classical PID is a linear feedback, see \cite{zhao2016,2022}. The boundedness of the partial derivatives $\frac{\partial g}{\partial x_i}$($i=1,2$) is equivalent to the global Lipschitz condition, which is a standard condition used in stochastic differential equation to ensure the global existence and uniqueness of solutions. In addition, the positive constants $L_1$, $L_2$, $N_1$, $N_2$,  $\underline b$ and $\bar b$ can be used to measure the system uncertainty quantitatively, which will also play a key role in designing the PID matrices.
\end{remark}

\noindent \textbf{Assumption 2.}
Both $r(t)$ and $d(t)$  are  continuously differentiable, and their time derivatives are  bounded, i.e.,
 $$\big|\dot r\big|_\infty:=\sup_{t\geq 0}|\dot r(t)|<\infty,~~~\big|\dot d\big|_{\infty}:=\sup_{t\geq 0}\big|\dot d(t)\big|<\infty.$$

\noindent\textbf{Assumption 3.}
The random noise along the reference signal is bounded, i.e.,
 $$|g(r,\dot r)|_\infty:=\sup_{t\geq 0}|g(r(t),\dot r(t))|<\infty.$$

\begin{remark}
Assumption 2 means that both the reference signal and the external disturbance cannot change ``too fast", since their $L^\infty$ norms limit the change rate. It is worth noting that the reference signal and the external disturbance are not required to be generated by some exosystems, which is a quite mild condition compared to those used in the literature for output regulation, see e.g., \cite{huang2004}.  Assumption 3 is also rather mild and reasonable, since the norm $|g(r,\dot r)|_\infty$ characterizes the \emph{intensity of random noise} along the reference signal.
Finally, we point out that the three norms $|\dot r|_\infty, |\dot d|_{\infty}$ and $|g(r,\dot r)|_{\infty}$ will directly affect the tracking performance of the PID controlled stochastic system, which can be seen from Theorem \ref{th1} given in the next section.
\end{remark}

\section{The main results}
\subsection{PID Parameter Design}
Stability is a primary requirement of a feedback system. It turns out that much insight into PID control can be obtained by analyzing the \emph{stability region}, which is the set of controller parameters that give stable closed-loop systems \cite{As}.

In this subsection, we will provide analytic design formulae for the PID matrices $K_p,K_i$ and $K_d$.

\textbf{Stability Region.} The PID gain matrices $K_p$, $K_i$ and $K_d$ are diagonal, positive definite and satisfy
	\begin{align}\label{th}\begin{split}
&\lambda_{\mathrm{min}}(K_i)>\lambda_{\mathrm{max}}(K_i)\big/\sqrt{2},\\
		&\lambda_{\mathrm{min}}\left(K_p^2-2K_iK_d\right)>\bar k+N_1^2\|K_d\|/\underline b,\\
& \lambda_{\mathrm{min}}\left(K_d^2-K_p/\underline b\right)>\bar k+N_2^2\|K_d\|/\underline b,\end{split}
	\end{align}
where $\bar k:=(L_1+L_2)(\|K_p\|+\|K_d\|)/\underline b$.

\begin{remark}\label{r1}
First, it is worth noting that inequalities (\ref{th}) for the PID matrices only involves the constants $L_1$, $L_2$, $N_1$, $N_2$ and $\underline b$, which are only related to properties of the system nonlinear functions $f$ and $g$. Therefore, the selection of the PID matrices is \emph{independent} of the reference signal $r(t)$ and the unknown disturbance $d(t)$. Next, we remark that inequalities (\ref{th}) always have solutions. Indeed, if the PID matrices are given by \begin{align}\label{pa}K_i=k_iI_n,~K_p=k_pI_n,~~K_d=k_d I_n,\end{align}
where $(k_i, k_p,k_d)\in\mathbb{R}^3$ are selected from the following  three dimensional unbounded set(\cite{Zhang2022}):
\begin{align}\label{pid}
\Omega=\left\{(k_i,k_p,k_d)\in\mathbb{R}^3_+ \left|
\begin{array}{c}
		~k_p^2-2k_ik_d>\bar k_1+k_dN_1^2/\underline b\\
		~k_d^2-k_p/\underline b>\bar k_1+k_dN_2^2/\underline b
	\end{array}\right.
\right\},
\end{align}
where $\bar k_1:=(L_1+L_2)(k_p+k_d)/\underline b$, then (\ref{th}) will be satisfied.
\end{remark}
\subsection{Bounded tracking error}
Let us consider the PID controlled stochastic system (\ref{sys})-(\ref{control}), where Assumptions 1-3 hold. Then we are in position to present the main results of this paper.
\begin{theorem}\label{th1} Suppose the PID matrices satisfy (\ref{th}), then there exist positive constants $c_1$, $c_2$ and $\lambda$,  depending at most on $K_i,K_p,K_d,L_1,L_2$, $N_1,N_2,\underline b,\bar b$ and $n$,  such that for all initial states $(x_1(0),x_2(0))\in\mathbb{R}^{2n}$,
	\begin{align}\label{3.4}\begin{split}
		\mathbf{E}\left(|e(t)|+|\dot e(t)|\right)^2~\leq&~ c_1\big(|e(0)|+|\dot e(0)|+|u^*|\big)^2e^{-\lambda t}\\
&+c_2\left(\big|\dot r\big|_\infty\!+\big|\dot d\big|_{\infty}+\big|g(r,\dot r)\big|_{\infty}\right)^2, ~\text{for all}~t\geq 0,\end{split}
	\end{align}
where $e(t)=r(t)-x_1(t)$ is the tracking error and $u^*$ meets the algebraic equation $f\left(r(0),\dot{r}(0),u^*\right)+d(0)=0$. In addition, if we have $\sup_{t\geq 0}|r(t)|+|d(t)|<\infty$, then the closed-loop system  (\ref{sys})-(\ref{control}) will be globally stable in the sense that $$\sup_{t\geq 0} \mathbf{E}\left(|x_1(t)|+|x_2(t)|+|u(t)|\right)^2<\infty.$$
\end{theorem}	
The proof of Theorem \ref{th1}  is supplied in the next section.

 We give some explanations of Theorem \ref{th1}.
Firstly, Theorem \ref{th1} demonstrates the ability of PID control to handle a vast range of uncertainties,
and ensures the \emph{boundedness} of the tracking error. Besides, by  (\ref{3.4}), it is not difficult to obtain
\begin{align}\label{upper}\begin{split}
\limsup_{t\to\infty} \mathbf{E}|e(t)|
		\le c\left(|\dot r|_\infty+\big|\dot d\big|_{\infty}+\big|g(r,\dot r)\big|_{\infty}\right),\end{split}
\end{align}
where the constant $c=\sqrt{c_2}$ relies on the choices of the PID matrices.
Inequality (\ref{upper}) indicates that the steady-state tracking error is determined by three critical factors:
$\!\!\!\!\!\!$\begin{itemize}
\item The quantity $|\dot r|_\infty+\big|\dot d\big|_\infty$, which reflects the \emph{change rate} of the reference signal $r(t)$ and the external disturbance $d(t)$;
\item The norm $|g(r,\dot r)|_{\infty}$, which characterizes the \emph{intensity of random noises} along the reference signal;
\item The constant $c$, which depends on the choices of the PID gain matrices.
\end{itemize}
In particular, one can see that the steady-state error will be \emph{small} if both $r(t)$, $d(t)$ are slowly time-varying and  the random noise is \emph{weak} along the reference signal.

Secondly, we remark that the boundedness of the reference signal and the external disturbance is \emph{not necessary} to obtain bounded tracking error, because we only require that the derivatives  $\dot r(t)$ and $\dot d(t)$ are bounded. Therefore, Theorem \ref{th1} has demonstrated the ability of PID to track a class of time-varying reference signals with a linear growth rate, which is a significant difference from most of the existing literature since the boundedness of  reference signals is usually required to obtain bounded tracking error, see e.g., \cite{guo2017,wang2022}.

Thirdly,  let us apply Theorem \ref{th1} to a special class of stochastic system, where the reference signal and the external disturbance  are \emph{constants}. The following corollary follows immediately from Theorem \ref{th1}.

\begin{corol}\label{coro1}
Consider the PID controlled system:
\begin{align*}
		&\mathrm{d}x_1=x_2 \mathrm{d}t,\\
		&\mathrm{d}x_2=[f(x_1,x_2,u)+d_0]\mathrm{d}t +g(x_1,x_2) \mathrm{d}B_t,~x_1,u\in\mathbb{R}^n,\\
&u=K_{p} e(t)+K_{i} \!\int_{0}^{t}\! e(s)\mathrm{d}s+K_{d} \dot{e}(t),
~~ e(t)=r^*-x_1(t),\end{align*}
where functions $f$ and $g$ satisfy Assumption 1,  $d_0\in\mathbb{R}^n$ is the external disturbance and $r^*\in\mathbb{R}^n$ is the setpoint.  Suppose $g(r^*,0)=0$ and PID matrices satisfy (\ref{th}), then there are some positive constants $c$ and $\lambda$, such that
	\begin{align*}
		\mathbf{E}\left(|e(t)|+|\dot e(t)|\right)^2\leq&~\! c\left(|e(0)|\!+\!|\dot e(0)|\!+\!|u^*|\right)^2e^{-\lambda t}, ~t\geq 0,
	\end{align*}
 where $u^*$ is the unique solution of $f\left(r^*,0,u\right)+d=0$.
\end{corol}
\begin{remark}From Corollary \ref{coro1}, one can see that for constant external disturbance and constant reference signal, the tracking error will converge to zero exponentially, provided that the diffusion term vanishes at $(r^*,0)$ and the PID matrices are chosen suitably. It is worth mentioning that Corollary  \ref{coro1} improves the corresponding results in \cite{Zhang2022} and \cite{2022}, in which the PID gains are limited to constants. Besides, Corollary \ref{coro1} also demonstrates that the PID controllers can deal with coupled non-affine  uncertain nonlinear systems, see  \cite{yuan2018} for related results of affine nonlinear systems without external disturbance.
\end{remark}

\subsection{Steady-state control performance}
\emph{Practical tracking} is an important research subject, which aims to drive the tracking error to a prescribed vicinity of zero in finite time and maintain it inside thereafter. From (\ref{upper}), one can see that the steady-state error depends on the choice of the PID matrices and the three $L^\infty$ norms $|\dot r|_\infty$, $|\dot d|_\infty$ and $|g(r,\dot r)|_\infty$.
However, it is still unclear whether PID control can be designed to achieve practical tracking, i.e., to make the steady-state error \emph{arbitrarily small}. In this subsection, we will address this basic question and give an affirmative answer to it.

To deal with this issue, we first point out an important geometric  property of the set $\Omega$ defined by (\ref{pid}), namely,
\begin{align}\label{101}
(k_i,k_p,k_d)\in \Omega \Rightarrow
\left(\frac{k_i}{\epsilon^3},\frac{k_p}{\epsilon^2},\frac{k_d}{\epsilon}\right)\in \Omega,~ 0<\epsilon\le 1.
\end{align}
Indeed, let $k_i^\epsilon=k_i/\epsilon^3,$ $k_p^\epsilon=k_p/\epsilon^2$, $k_d^\epsilon=k_d/\epsilon,$ and $\bar k_1^\epsilon=(k_p^\epsilon+k_d^\epsilon)(L_1+L_2)$. Note that $0<\epsilon\le 1$, it is easy to see
\begin{align*}
&(k_p^\epsilon)^2-2k_i^\epsilon k_d^\epsilon=\epsilon^{-4} (k_p^2-2k_ik_d)>\epsilon^{-4} (\bar k_1+k_dN_1^2/\underline b)
\geq \bar k_1^\epsilon +k_d^\epsilon N_1^2/\underline b,\\
&(k_d^\epsilon )^2-k_p^\epsilon/\underline b =\epsilon^{-2}(k_d^2-k_p/\underline b)>\epsilon^{-2}(\bar k_1+k_dN_2^2/\underline b)\geq \bar k_1^\epsilon+k_d^\epsilon N_2^2/\underline b,
\end{align*}
thus $(k_i^\epsilon,k_p^\epsilon,k_d^\epsilon)\in \Omega$.

We also need an additional assumption of the reference signal.

\noindent \textbf{Assumption 4.}
 The reference signal $r(\cdot)\in C^2(\mathbb{R}^+,\mathbb{R}^n)$, and its second derivative is bounded on $\mathbb{R}^+$.

Now, consider the PID controlled stochastic system (\ref{sys})-(\ref{control}), where Assumptions 1-4 hold. By virtue of property (\ref{101}), we have the following result.
\begin{theorem}\label{th2} Let $(k_i,k_p,k_d)\in\Omega$ be any given triple.  Suppose the PID gain matrices are taken as follows:
	\begin{align}\label{parameter}		K_i^\epsilon=\frac{k_iI_n}{\epsilon^3},~K_p^\epsilon=\frac{k_pI_n}{\epsilon^2},~K_d^\epsilon=\frac{k_dI_n}{\epsilon},~ 0<\epsilon\le 1.
	\end{align}
	Then there exists a constant $c>0$, which  depends on $k_i,k_p,k_d,L_1,L_2,N_1,N_2,\underline b,\bar b,n$ only, such that the following tracking performance will be achieved:
	\begin{align}\label{c}\begin{split}
		\limsup_{t\to\infty} \mathbf{E}|e(t)|^2&
\le c\left[(|\dot r|_{\infty}+|\ddot r|_{\infty}+|\dot d|_{\infty})^2\epsilon+|g(r,\dot r)|_\infty^2\right]\epsilon^3\\
\limsup_{t\to\infty} \mathbf{E}|\dot e(t)|^2&\le
c\left[(|\dot r|_{\infty}+|\ddot r|_{\infty}+|\dot d|_{\infty})^2\epsilon+|g(r,\dot r)|_\infty^2\right]\epsilon\end{split}
	\end{align}
for all $0<\epsilon\le 1$, where $|\dot r|_\infty$, $|\ddot r|_\infty$, $|\dot d|_{\infty}$ and $|g(r,\dot r)|_\infty$ are the $L^\infty$ norms of the corresponding functions.
\end{theorem}

\begin{remark}We remark that the constant $c$ in Eq. (\ref{c}) is independent of the parameter $\epsilon$ and these $L^\infty$ norms. Therefore, practical tracking can be achieved for the nonlinear stochastic model (\ref{sys}), as long as one tunes $\epsilon$ \emph{small enough}.
In fact, it follows from Theorem \ref{th2} that the steady-state error will be small, provided that one of the following conditions holds:
\begin{itemize}
\item The norms $|\dot r|_\infty$, $|\ddot r|_\infty$, $|\dot d|_{\infty}$, $|g(r,\dot r)|_\infty$ are small;
\item The PID gains are large enough according to (\ref{parameter}).
\end{itemize}
\end{remark}

\subsection{Transient control performance}
In practical applications, the quality of control systems is usually evaluated by both the steady-state and transient performance. Theorem \ref{th2} indicates that nice steady-state control performance can be guaranteed by choosing sufficiently large PID parameters. Of course, it is natural to ask whether PID controllers can be designed to ensure both nice transient and steady errors during the tracking process.

For this purpose, we introduce the following desired transient process to be tracked by $x_1(t)$, which is shaped from the reference signal $r(t)$ by a  stable linear filter(\cite{zhong}):
\begin{align}\label{filter}\begin{split}
&\ddot y^*=-\omega^2(y^*-r(t))-2\omega(\dot y^*-\dot r(t))+\ddot r(t),\\
&y^*(0)=x_1(0),~~\dot y^*(0)=x_2(0),\end{split}
\end{align}
where $\omega>0$ is a constant for tuning the speed of the transient process, and $x_1(0),x_2(0)$ are the initial states of the stochastic system (\ref{sys}).

Now, suppose the PID control for system (\ref{sys}) is described as follows:
\begin{align}\label{pE}\begin{split}
\hat u(t)=&K_p E(t)+K_i\int_0^t E(s)ds+K_d \dot{E}(t),
\end{split}
\end{align}
where $E(t)=y^*(t)-x_1(t)$ is the error between the controlled variable $x_1(t)$ and its ideal trajectories $y^*(t)$.

To ensure desired transient performance, the controlled variable $x_1(t)$ should be close to $y^*(t)$ during the tracking process. This inspires us to consider the following bound:
\begin{align}
\sup_{t\geq 0} \mathbf{E}\left[\left|x_1(t)-y^*(t)\right|^2+\left|x_2(t)-\dot{y}^*(t)\right|^2\right].
\end{align}
Next, we will show that one can design PID matrices such that the above upper bound is as small as possible.
\begin{theorem}\label{th3}
Consider system (\ref{sys}),  where Assumptions 1-4 hold and the control $\hat u$ is defined by (\ref{pE}).  Suppose PID matrices are given by (\ref{parameter})  with $(k_i,k_p,k_d)\in\Omega$ be a given triple. Then there is a constant $c>0$, which is independent of $\epsilon$, such that the following estimates of the tracking errors hold uniformly in $[0,\infty)$, i.e.,
\begin{align}\begin{split}
\sup_{t\geq 0}\mathbf{E}\left|E(t)\right|^2\le c\epsilon^3,~
\sup_{t\geq 0}\mathbf{E}\left|\dot {E }(t)\right|^2\le c\epsilon, ~0<\epsilon\le 1,\end{split}
\end{align}
where $E(t)=y^*(t)-x_1(t)$.
\end{theorem}
\begin{remark}
First, Theorem \ref{th3} shows that the controlled variable $x_1(t)$ can be tuned arbitrarily close to its ideal transient process $y^*(t)$,  provided that one increases the PID gains suitably according to formula (\ref{parameter}). Next,
from the proof of Theorem \ref{th3}, it can be obtained that
\begin{align*}
\sup_{t\geq 0}\mathbf{E}|u(t)|^2=O(\epsilon^{-1}), ~0<\epsilon\le 1,
\end{align*}
which implies that the amplitude of controller $u(t)$ is of order $1/\sqrt{\epsilon}$ as $\epsilon\to 0$.
Therefore, large PID gains(i.e., $\epsilon$ is chosen sufficiently small) will make the implementation of the controller impossible, since saturation of physical actuators is ubiquitous in practical control systems. Besides, large PID gains may amplify measurement errors due to the inevitability of measurement noises, which may lead to unsatisfactory tracking performance. Therefore, the theoretical analysis on PID controlled nonlinear stochastic systems with measurement noises and input saturation should be an important work in the future.
\end{remark}

\section{Proofs of the main results}
To prove the main results, we need to introduce several lemmas.

First, define a function space $C_b^k\left([0,\infty),\mathbb{R}^n\right)$, which consists of all functions $h\in C^k\big([0,\infty),\mathbb{R}^n\big)$ with property
$$\sup_{t\geq 0}\left|\dot h(t)\right|+\cdots+ \left|h^{(k)}(t)\right|<\infty, $$
where $k\geq 1$ is a positive integer.

The first lemma indicates that any function $h\in C_b^1([0,\infty),\mathbb{R}^n)$ can be approximated by some function  $p\in C_b^3([0,\infty),\mathbb{R}^n)$ in a certain sense. To be specific,
\begin{lemma}\label{lem1}
For any given $n\geq 1$, there is a constant $c_n>0$, such that for all $h\in C_b^1([0,\infty),\mathbb{R}^n)$, there exists some $p\in C_b^3([0,\infty),\mathbb{R}^n)$, which satisfies
\begin{align}\label{distance}
\begin{split}
&|p(t)-h(t)|\vee \left|\dot p(t)\right|\vee \left |\ddot p(t)\right|\vee\left|p^{(3)}(t)\right|
\le ~c_n\sup_{t\geq 0}\left|\dot h(t)\right|,~\text{for all}~~t\geq 0.\end{split}
\end{align}
\end{lemma}

\begin{lemma} \label{lem2}Suppose that the $n\times n$  matrices $K_p$, $K_i$ and $K_d$ are  diagonal, positive definite and satisfy inequalities (\ref{th}), then the following $3n\times 3n$ matrix
\begin{align}\label{P}
	P\overset{\triangle}{=}\frac{1}{2}
	\begin{bmatrix}
		2\underline bK_iK_p&2\underline bK_iK_d&K_i\\
		2\underline bK_iK_d&~2\underline bK_pK_d-K_i~&K_p\\
		K_i&K_p&K_d
	\end{bmatrix}
\end{align}
is positive definite. Also, let $A,B,C,D$ and $\theta$ be  $n\times n$ matrices that satisfy
$$\|A\|\le L_1,~\|B\|\le L_2,~\|C\|\le N_1,~\|D\|\le N_2,~\theta\geq \underline b I_n.$$
 Then,  there exists $\gamma>0$, which depends on $K_p$, $K_i$, $K_d$, $L_1$, $L_2$, $ N_1$, $N_2$ and $\underline b$ only, such that
  for all $Y=[y_0^\top,y_1^\top,y_2^\top]^\top\in\mathbb{R}^{3n}$,
\begin{align*}
2Y^\top P\begin{bmatrix}y_1\\y_2\\Ay_1+By_2-\theta\bar y\end{bmatrix}\!+\frac{1}{2}\|K_d\||Cy_1+D y_2|^2
\le -\gamma |Y|^2,
\end{align*}
where $\bar y:=K_iy_0+K_py_1+K_d y_2$.
\end{lemma}

The proofs of Lemmas \ref{lem1}-\ref{lem2} are given in the Appendix.

It is noteworthy that the positive constant $\gamma$ is independent \emph{of the five matrices $A,B,C,D$ and $\theta$}. This property is crucial in the proofs of the main results.

\begin{lemma}\cite[Proposition 4.1]{2022}\label{lem3}
Let $\Phi \in C^1(\mathbb{R}^n,\mathbb{R}^n)$. Suppose that  \begin{align*}\frac{\partial \Phi}{\partial x}\geq R,~\forall x\in\mathbb{R}^n,\end{align*}
where $R\in\mathbb{R}^{n\times n}$ is a constant positive definite matrix,  then $\Phi$ is a global diffeomorphism on $\mathbb{R}^n$. As a consequence, $\Phi$ is surjective.
\end{lemma}

We also need some basic concepts and results about the solution of stochastic differential equations(SDE).
Consider the following nonlinear SDE:
\begin{equation}
	\begin{cases}
		\mathrm{d}x(t)\!\!& =b(t,x(t))\mathrm{d}t+\sigma(t,x(t))\mathrm{d}B_t,\\
		x(0)\!\!& =x_{0},
	\end{cases}\label{eq:sto}
\end{equation}
where $x\in\mathbb{R}^{m}$ is the state, $B_t$ is a one-dimensional
standard Brownian motion defined on a complete probability space $(\Omega,\mathscr{F},P)$
with $\{\mathscr{F}_{t}\}_{t\geq 0}$ being a natural filtration, and
$b(t,x)\in C^{1}(\mathbb{R}^+\times\mathbb{R}^{m},\mathbb{R}^{m}),$ $\sigma(t,x)\in C^{1}(\mathbb{R}^+\times\mathbb{R}^{m},\mathbb{R}^{m})$ are nonlinear functions.

\begin{definition}\label{de1}
Given a time-invariant function $V\in C^{2}(\mathbb{R}^{m},\mathbb{R})$ associated with the equation (\ref{eq:sto}). The differential operator $\mathcal{L}$ acting on $V$ is defined by
	\begin{align}\label{definition of operator L}
\mathcal{L}V(t,x)=\frac{\partial V}{\partial x}b(t,x) +\frac{1}{2}\sigma(t,x)^{\top}\frac{\partial^{2}V}{\partial x^{2}}\sigma(t,x),
	\end{align}
where $\frac{\partial V}{\partial x}:=(\frac{\partial V}{\partial x_1},\cdots,\frac{\partial V}{\partial x_m} )$ and $\frac{\partial^{2}V}{\partial x^{2}}:=(\frac{\partial^2 V}{\partial x_i\partial x_j})_{m\times m}$ are the gradient and Hessian matrix of $V$ respectively.
\end{definition}

According to  \cite[Theorem 3.5 and Remark 3.4]{Khasminskii} up to a slight modification, we have the following lemma.
\begin{lemma}\label{lem4}
	If there exist a nonnegative function $V\in C^{2}(\mathbb{R}^n,\mathbb{R})$ and positive constants $\alpha,\beta$ such that $\liminf_{|x|\to \infty}V(x)=\infty$ and $\mathcal{L}V(t,x)\leq -\alpha V(x)+\beta$,
	then SDE \eqref{eq:sto} has a unique solution $(x(t))_{t\geq 0}$ such that
	\begin{equation*}
		\mathbf{E}V(x(t))\leq V(x(0))e^{-\alpha t}+\beta/\alpha,~\text{for all}~t\geq 0.
	\end{equation*}
\end{lemma}
\textbf{Proof of Theorem \ref{th1}.}	
For simplicity, let us denote
$$M\!:=|\dot r|_\infty, ~M_1\!:=\big|\dot d\big|_\infty,~M_2\!:=|g(r,\dot r)|_\infty.$$

\emph{Step 1.} First, by applying Lemma \ref{lem1}  to $r(t)$, there exists $q(t)\in C_b^3([0,\infty), \mathbb{R}^n)$ such that
\begin{align}\label{r}
\max\left\{|q(t)-r(t)|, |\dot q(t)|,  \big|\ddot q(t)\big|\right\} \le c_n M,~\forall t\geq 0,
\end{align}
where $c_n>0$ only depends on $n$.
Next, by condition $ \frac{\partial f}{\partial u}\geq \underline b I_n$ and Lemma \ref{lem3},
 we know that for any given $t\geq 0$, there exists a unique $u_r(t)\in\mathbb{R}^n$, such that
 \begin{align}\label{identity}
	f\big(r(t),\dot r(t),u_r(t)\big)+d(t)=0.
\end{align}
Similarly, for all $t\geq 0$,  there exists a unique $u_q(t)\in\mathbb{R}^n$, such that
 \begin{align}\label{identity1}
	f\big(q(t),\dot q(t),u_q(t)\big)+d(t)=0.
\end{align}
Combine (\ref{identity}) with (\ref{identity1}), we obtain
\begin{align}\label{id}\begin{split}
&f(r(t),\dot r(t),u_r(t))-f(r(t),\dot r(t), u_q(t))\\
=&f(q(t),\dot q(t),u_q(t))-f(r(t),\dot r(t),u_q(t)).\end{split}
\end{align}
Note that $0<\underline b I_n\le \frac{\partial f}{\partial u}\le \bar b I_n$,  for any given $t\geq 0$, it can be derived that there exists some $n\times n$ matrix $\theta_t$, such that
\begin{align}\label{ie}\begin{split}
&f(r(t),\dot r(t),u_r(t))-f(r(t),\dot r(t), u_q(t))\\
=&\theta_t(u_r(t)-u_q(t)),~~~0<\underline bI_n\le \theta_t\le \bar b I_n.\end{split}
\end{align}
Therefore, it follows from (\ref{id}), (\ref{ie}) and (\ref{r}) that
\begin{align}\label{ur}
|u_r(t)-u_q(t)|=&
\left|\theta_t^{-1}\left [f(r(t),\dot r(t),u_r(t))-f(r(t),\dot r(t), u_q(t))\right]\right|\nonumber\\
=&\left|\theta_t^{-1}\left[f(q(t),\dot q(t),u_q(t))-f(r(t),\dot r(t),u_q(t))\right]\right|\nonumber\\
\le&~\! \frac{1}{\underline b}\Big[L_1\sup_{t\geq 0}|q(t)-r(t)|~\!+~\!L_2\sup_{t\geq 0} \left|\dot q(t)-\dot r(t)\right|\Big]\nonumber\\
\le& ~\!\frac{1}{\underline b}\Big[L_1\sup_{t\geq 0}|q(t)-r(t)|+\!L_2\sup_{t\geq 0} \left|\dot q(t)\right|+\left|\dot r(t)\right|\Big]\nonumber\\
\le & ~\!\left[(L_1c_n+L_2(c_n+1))/\underline b \right]M,~\text{for all~} t\geq 0.
\end{align}
 On the other hand, by taking the derivative of both sides of (\ref{identity1}), it follows that
\begin{align*}
\frac{\partial f}{\partial q}\dot q(t)+\frac{\partial f}{\partial \dot q}\ddot q(t)+\frac{\partial f}{\partial u_q}\dot u_q(t)+\dot d(t)=0.
\end{align*}
Hence, by (\ref{r}) and Assumption 1, we obtain the following estimate:
\begin{align}\label{uq}
\left|\dot u_q(t)\right|=&
\left|\left[\frac{\partial f}{\partial u_q}\right]^{-1}\left(\frac{\partial f}{\partial q}\dot q(t)+\frac{\partial f}{\partial \dot q}\ddot q(t)+\dot d(t)\right)\right|\nonumber\\
\le &\frac{1}{\underline b}\left[c_n(L_1+L_2)M+M_1\right]\nonumber\\
\le &\frac{1}{\underline b}\left[c_n(L_1+L_2)+1\right](M+M_1).
\end{align}
From (\ref{uq}), one can see Lemma  \ref{lem1} can be applied to the function $u_q$. Therefore, there exist some $v_q\in C^3_b([0,\infty),\mathbb{R}^n)$ such that
\begin{align}\label{vq}
|v_q(t)-u_q(t)|\vee|\dot v_q(t)|\vee |\ddot v_q(t)|\vee \big|v_q^{(3)}(t)\big| \le \eta_0(M+M_1),
\end{align}
where $\eta_0:=(c_n^2(L_1+L_2)+c_n)/\underline b$ depends on $L_1,L_2,\underline b$ and $n$ only. Combine (\ref{ur}) and (\ref{vq}), it follows that
\begin{align}\label{urvq}
&|v_q(t)-u_r(t)|\le |v_q(t)-u_q(t)|+|u_q(t)-u_r(t)|\nonumber\\
\le &\left[ \eta_0+(L_1c_n+L_2(c_n+1))/\underline b\right](M+M_1)\nonumber\\
=:&\eta_1(M+M_1).
\end{align}
The constant $\eta_1$ also depends on $L_1,L_2,\underline b$ and $n$ only.

\emph{Step 2.} Based on $q(t)$ and $v_q(t)$ given in Step 1, we introduce some notations
\begin{align}
	y_0(t):=&-\int_0^t e(s)\mathrm{d}s+K_i^{-1}v_q(t),\label{y0}\\
\label{y1}
\begin{split}
	y_1(t):=&~x_1(t)-q(t)+K_i^{-1}\dot v_q(t)\\
=&-e(t)+\left[ r(t)-q(t)+K_i^{-1}\dot v_q(t)\right],
\end{split}\\
\label{y2}
\begin{split}
y_2(t):=&~x_2(t)-\dot q(t)+K_i^{-1}\ddot v_q(t)\\
=&-\dot e(t)+\left[\dot r(t)-\dot q(t)+K_i^{-1}\ddot v_q(t)\right],\end{split}\\
	\bar y(t):=&~\!K_iy_0(t)+K_py_1(t)+K_dy_2(t),
\end{align}
then the PID control (\ref{control}) can be written as follows:
\begin{align*}
	u(t)=&~K_{p} e(t)+K_{i}\int_{0}^{t} e(s)\mathrm{d}s+K_{d} \dot{e}(t)\nonumber\\
=&-\bar y(t)+v_q(t)+K_p\left[r(t)-q(t)+K_i^{-1}\dot v_q(t)\right]\nonumber\\
&+K_d\left[\dot r(t)-\dot q(t)+K_i^{-1}\ddot v_q(t)\right].
\end{align*}
For simplicity, let us denote
\begin{align}\label{tri1}
\triangle_1(t):=&v_q(t)-u_r(t)+K_p\left[r(t)-q(t)+K_i^{-1}\dot v_q(t)\right]\nonumber\\
&+K_d\left[\dot r(t)-\dot q(t)+K_i^{-1}\ddot v_q(t)\right],
\end{align}
then  $u(t)$ has a simpler expression:
\begin{align}\label{u}
	u(t)=&-\bar y(t)+\triangle_1(t)+u_r(t).
\end{align}
Now, we proceed to estimate the upper bound of $\triangle_1(t)$.
By definition (\ref{tri1}) of $\triangle_1(t)$, it is easy to see
\begin{align}
|\triangle_1(t)|\le& |v_q(t)-u_r(t)|+\|K_p\|\left(|r(t)-q(t)|+|K_i^{-1}\dot v_q(t)|\right)\nonumber\\
&+\|K_d\|\left(|\dot r(t)|+|\dot q(t)|+|K_i^{-1}\ddot v_q(t)|\right).
\end{align}
From (\ref{r}), (\ref{vq}) and (\ref{urvq}), there exists constant $\eta_2>0$, which depends on $L_1,L_2,\underline b$, $K_p,K_i,K_d$ and $n$ only, such that
 \begin{align}\label{tri 11}|\triangle_2(t)|\le \eta_2(M+M_1), ~\text{for all~}t\geq 0.\end{align}

\emph{Step 3.} By definitions (\ref{y0})-(\ref{y2}), we obtain the following SDE:
\begin{align}\label{sde1}\begin{split}
\mathrm{d}y_0=&\left[y_1+(q(t)-r(t))\right]\mathrm{d}t\\
\mathrm{d}y_1=&~y_2\mathrm{d}t\\
\mathrm{d}y_2=&~\mathrm{d}x_2+\left(K_i^{-1}v_q^{(3)}(t)-\ddot q(t)\right)\mathrm{d}t.\end{split}
\end{align}
Besides, recall $\mathrm{d}x_2(t)=[f(x_1(t),x_2(t),u(t))+d(t)]\mathrm{d}t+g(x_1(t),x_2(t))\mathrm{d}B_t$, and apply relationship (\ref{identity}) and Assumption 1, we have
\begin{align}\label{dx2}
&f(x_1(t),x_2(t),u(t))+d(t)\nonumber\\=&f(x_1(t),x_2(t),u(t))-f(r(t),\dot r(t),u_r(t))\nonumber\\
=&A(x_1(t)-r(t))+B(x_2(t)-\dot r(t))+\theta (u(t)-u_r(t))
\end{align}
for some time-varying $n\times n$  matrices  $A$, $B$, $\theta$ satisfying
$$\|A\|\leq L_1, ~~\|B\|\leq L_2,~~
 \underline b I_n\le \theta \le  \bar b I_n.$$
 Similarly, it can be obtained that
 \begin{align}\label{dx22}
&g(x_1(t),x_2(t))-g(r(t),\dot r(t))
=C(x_1(t)-r(t))+D(x_2(t)-\dot r(t))
\end{align}
for some time-varying $n\times n$ matrices $C$ and $D$ satisfying
$$\|C\|\leq N_1, ~~\|D\|\leq N_2.$$
Next, from  (\ref{y1}), (\ref{y2}) and  (\ref{u}), we know
$$x_1(t)-r(t)=y_1(t)+\triangle_2(t), ~x_2(t)-\dot r(t)=y_2(t)+\triangle_3(t),~ u(t)-u_r(t)=-\bar y(t)+\triangle_1(t),$$
where $\triangle_2(t)$ and $\triangle_3(t)$ are time-varying functions defined by
\begin{align}\begin{split}
\triangle_2(t):=q(t)-r(t)-K_i^{-1}\dot v_q(t),\\
\triangle_3(t):=\dot q(t)-\dot r(t)-K_i^{-1}\ddot v_q(t).\end{split}
\end{align}
Therefore, we have
\begin{align}\label{f}\begin{split}
&f(x_1(t),x_2(t),u(t))+d(t)\\
=&A(x_1(t)-r(t))+B(x_2(t)-\dot r(t))+\theta (u(t)-u_r(t))\\
=&A(y_1(t)+\triangle_2(t))+B(y_2(t)+\triangle_3(t))+\theta (-\bar y(t)+\triangle_1(t))\\
&g(x_1(t),x_2(t))=C(x_1(t)-r(t))+D(x_2(t)-\dot r(t))+g(r(t),\dot r(t))\\
=&C(y_1(t)+\triangle_2(t))+D(y_2(t)+\triangle_3(t))+g(r(t),\dot r(t)).\end{split}
\end{align}
Let us define
\begin{align}\label{tri5}\begin{split}
\triangle_4(t):=&A\triangle_2(t)+B\triangle _3(t)+\theta\triangle_1(t)+K_i^{-1}v_q^{(3)}(t)-\ddot q(t),\\
\triangle_5(t):=&C\triangle_2(t)+D\triangle_3(t)+g(r(t),\dot r(t)),\end{split}
\end{align}
then from (\ref{sde1}), (\ref{f}) and (\ref{tri5}), we have
\begin{align*}
\begin{split}
\mathrm{d} y_0(t)=&\left[y_1(t)+(q(t)-r(t))\right]\mathrm{d}t\\
 \mathrm{d}y_1(t)=&~\!y_2(t)\mathrm{d}t\\
\mathrm{d}y_2(t)=&\left[(Ay_1(t)+By_2(t)-\theta \bar y(t))+\triangle_4(t)\right]\mathrm{d}t
+\left[Cy_1(t)+Dy_2(t)+\triangle_5(t)\right]\mathrm{d}B_t.\end{split}
\end{align*}
For simplicity, set
\begin{align*}
Y=&~\begin{bmatrix}~y_0~\\~y_1~\\~y_2~\end{bmatrix},~~~~~
b(t,Y)=\begin{bmatrix}y_1\\y_2\\Ay_1+By_2-\theta\bar y
\end{bmatrix},\\\\
\triangle=&\begin{bmatrix}q-r\\\mathbf{0}\\\triangle_4\end{bmatrix},
~~~~\sigma(t,Y)=\begin{bmatrix}\mathbf{0}\\\mathbf{0}\\Cy_1+Dy_2+\triangle_5
\end{bmatrix},
\end{align*}
where $\mathbf{0}$ is the $n\times 1$  vector with zero entries. Then we have
\begin{align}\label{SDE}
\begin{split}
\mathrm{d}Y(t)=
\left[b(t,Y(t))+\triangle(t)\right] \mathrm{d}t+\sigma(t,Y(t)) \mathrm{d}B_t.
\end{split}
\end{align}
By definition (\ref{tri5}), one can easily see that there exists some constant $\eta_3>0$, which only depends  on
 $K_p,K_i,K_d,L_1,L_2$, $N_1$, $N_2$, $\underline b$, $\bar b$ and $n$, such that
\begin{align}\label{tri15}|\triangle(t)|\le \eta_3(M+M_1),~~~ |\triangle_5(t)|\le \eta_3(M+M_1) +M_2.\end{align}
\emph{Step 4.} Inspired by \cite{2022}, we will consider the Lyapunov function $V(Y)=Y^\mathrm{T} PY$, where $P$ is a $3n\times 3n$ matrix defined by (\ref{P}).
From Lemma \ref{lem2}, the matrix $P$ is positive definite. Moreover, by Definition \ref{de1},  the differential operator $\mathcal{L}$ acting on $V$ is given by
\begin{align*}
\mathcal{L} V=&\underbrace{\frac{\partial V}{\partial Y}b(t,Y)}_{\mathrm{I}}+
\underbrace{\frac{\partial V}{\partial Y}\triangle(t)}_{\mathrm{II}} +
\underbrace{\sigma(t,Y)^{\top}P\sigma(t,Y)}_{\mathrm{III}}.
\end{align*}
By some simple calculations, it can be deduced that
\begin{align*}
&\mathrm{I}=~2Y^{\top}P b(t,Y),~~\mathrm{II}=~\!2Y^\top P\triangle (t),\\
&\mathrm{III}=~\!\frac{1}{2}(C y_1+D y_2 +\triangle_5(t))^{\top}K_d(C y_1+Dy_2 +\triangle_5(t)).
\end{align*}

\emph{Step 5.} We next estimate the upper bound of $\mathcal{L} V.$
From (\ref{tri15}), the second term in $\mathcal{L} V$ satisfies
\begin{align}\label{ii}
|\mathrm{II}|\le 2\|P\||Y| |\triangle|\le \frac{2\|P\|\eta_3(M+M_1)\sqrt{V}}{\sqrt{\lambda_{\text{min}}(P)}}.
\end{align}
Besides, recall $|\triangle_5(t)|\le \eta_3(M+M_1) +M_2$, $\|C\|\le N_1,~\|D\|\le N_2$, thus the third term  has the following upper bound:
\begin{align}\label{iiii}\begin{split}
\mathrm{III}=&~ \!\frac{1}{2}\left[(Cy_1+Dy_2)^{\top}K_d(Cy_1+Dy_2)+\triangle_5^{\top}K_d \triangle_5\right]
+(Cy_1+Dy_2)^{\top}K_d \triangle_5\\
\le &~\!\frac{1}{2}\|K_d\||C y_1+Dy_2|^2+\|K_d\|(\eta_3^2(M+M_1)^2\!+M_2^2)\\
&+(N_1+N_2)\|K_d\|(\eta_3(M+M_1)+M_2)|Y|.\end{split}
\end{align}
Combine (\ref{ii}) and (\ref{iiii}), we conclude that
\begin{align*}
\mathcal{L}V~\le  &~~~2Y^{\top}P b(t,Y)+\frac{1}{2}\|K_d\||C y_1+D y_2|^2\\
&+\eta_4\left[(M+M_1+M_2)^2+(M+M_1+M_2)\sqrt{V}\right]
\end{align*}
for some  $\eta_4$ depending on $K_i,K_p,K_d,L_i,N_i,\underline b,\bar b,n$ only.
Moreover, by Lemma \ref{lem2}, there exists some positive constant $\lambda$ which depends on $(K_i,K_p,K_d,L_i,N_i,\underline b)$ only, such that $2Y^{\top}P b(t,Y)+\frac{1}{2}\|K_d\||C y_1+D y_2|^2\le -2\lambda V$, which gives
\begin{align*}
\mathcal{L}V\le  &-2\lambda V+\eta_4\left[\hat M^2+\hat M \sqrt{V}\right]\\
\le &-\lambda V+\left(\eta_4+\eta_4^2/(4\lambda)\right)\hat M^2,
\end{align*}
where $\hat M:=M+M_1+M_2.$
Therefore, by Lemma \ref{lem4}, we know that SDE \eqref{SDE} has a unique solution $(Y(t))_{t\geq 0}$ such that
$\mathbf{E}V(t)\leq V(0)e^{-\lambda t}+\left(\eta_4+\eta_4^2/(4\lambda)\right)\hat M^2/\lambda.$	
 As a consequence, we have
\begin{align*}
\mathbf{E}\sum_{i=0}^2|y_i(t)|^2
\le \eta_5\bigg[\sum_{j=0}^2|y_j(0)|^2e^{-\lambda t}+\hat M^2\bigg],
\end{align*}
where $$\eta_5:=\max\{2(\eta_4+\eta_4^2/\lambda)/\lambda,\lambda_{\text{max}}(P)\}/\lambda_{\text{min}}(P)$$ depends on $(K_i,K_p,K_d,L_i,N_i,\underline b,\bar b,n)$ only.
It follows from (\ref{y1})-(\ref{y2}) that
$$e(t)=-y_1(t)+\left(r(t)-q(t)+K_i^{-1}\dot v_q(t)\right),~~\dot e(t)=-y_2(t)+\left(\dot r(t)-\dot q(t)+K_i^{-1}\ddot v_q(t)\right),$$
therefore it can be obtained
\begin{align*}
&|e(t)|^2\le2|y_1(t)|^2+ \eta_{6}(M+M_1)^2,~~|\dot e(t)|^2\le 2|y_2(t)|^2+\eta_{6} (M+M_1)^2,\\
&|y_1(0)|^2\le 2|e(0)|^2+\eta_{6} (M+M_1)^2,~~|y_2(0)|^2\le 2|\dot e(0)|^2+\eta_{6} (M+M_1)^2,
\end{align*}
for some $\eta_6$ depending on $(K_i,K_p,K_d,L_i,N_i,\underline b,\bar b,n)$ at most.
Note also that $y_0(0)=K_i^{-1}v_q(0)$, thus there exists $\eta_7>0$ such that
\begin{align*}|y_0(0)|^2\le& 2\left[|K_i^{-1}(v_q(0)-u_r(0))|^2+|K_i^{-1}u_r(0)|^2\right]\\
\le &\eta_7((M+M_1)^2+|u_r(0)|^2).\end{align*}
Consequently, we have
\begin{align*}
&\mathbf{E}\left[|e(t)|^2+|\dot e(t)|^2\right]\\
\le& 2\mathbf{E}\left[|y_1(t)|^2+|y_2(t)|^2+\eta_{6}(M+M_1)^2 \right]\nonumber\\
\le & ~\!\eta_{8}\bigg(\sum_{i=0}^2|y_i(0)|^2e^{-\lambda t}+\hat M^2\bigg)\nonumber\\
\le&  c_1\left(|e(0)|^2+|\dot e(0)|^2+|u_r(0)|^2\right)e^{-\lambda t}+c_2\hat M^2,~ t\geq 0
\end{align*}
for some $c_1$ and $c_2$ depending on $K_i,K_p,K_d,L_i,N_i,\underline b,\bar b$ and $n$ only. Note that $u_r(0)$ satisfies the equation $f(r(0),\dot r(0), u_r(0))+d(0)=0$. Recall $\hat M=M+M_1+M_2$, therefore the first statement of Theorem \ref{th1} is proved.

Finally, if in addition we have the condition that  $r$ and $d$ are bounded, then from (\ref{y1}), (\ref{y2}) and (\ref{u}), it follows that
\begin{align}\label{bound}\begin{split}
&|x_1(t)|\le |y_1(t)|+\underbrace{|q(t)-r(t)|+|r(t)|+|K_i^{-1}\dot v_q(t)|}_{\text{bounded}},\\
&|x_2(t)|\le |y_2(t)|+\underbrace{|\dot q(t)|+|K_i^{-1}\ddot v_q(t)|}_{\text{bounded}},\\
&|u(t)|\le |\bar y(t)|+\underbrace{|\triangle_1(t)|+|u_r(t)|}_{\text{bounded}}.\end{split}
\end{align}
Thus, property $\sup_{t\geq 0} \mathbf{E}\left(|x_1(t)|+|x_2(t)|+|u(t)|\right)^2<\infty$
follows immediately from (\ref{bound}) and the boundedness of $\mathbf{E}\sum_{i=0}^2|y_i(t)|^2$.
\hfill $\square$

\begin{remark}
It is worth mentioning that the positive constants $\eta_1, \cdots,\eta_8$, $c_1$, $c_2$ and $\lambda$ that appeared in the proof of Theorem \ref{th1}  are independent of the initial states, the reference signal and the external disturbance, since they depend at most on the constants $L_1,L_2$, $N_1,N_2,\underline b,\bar b,n$ and the PID parameters.
\end{remark}

\textbf{Proof of Theorem \ref{th2}.}
Now, suppose $K_p$, $K_i$ and $K_d$ are determined by formula (\ref{parameter}). Let us denote
\begin{align}
&w_0(t):=-\frac{1}{\epsilon^2}\int_0^t e(s)\mathrm{d}s+\frac{\epsilon }{k_i}v_q(t),\label{int} \\ &w_1(t):=-\frac{1}{\epsilon}e(t)+\frac{\epsilon^2 }{k_i}\dot v_q(t),\label{pe}\\
&w_2(t):=-\dot e(t)+\frac{\epsilon^3}{k_i}\ddot v_q(t),\label{pd}\\
&\bar w(t):= k_iw_0(t)+k_pw_1(t)+k_dw_2(t),\label{bar y}
\end{align}
where the function $v_q\in C^3_b([0,\infty),\mathbb{R}^n)$ is given in Step 1 in the proof of Theorem \ref{th1},
then the PID control can be written as follows:
\begin{align}
u(t)=-\frac{\bar w(t)}{\epsilon}+u_r(t)+(v_q(t)-u_r(t))+\frac{k_p}{k_i}\epsilon\dot v_q(t)+\frac{k_d}{k_i}\epsilon^2\ddot v_q(t).
\end{align}
Moreover, denote
\begin{align}\begin{split}
&\delta_1(t):=v_q(t)-u_r(t)+\frac{k_p}{k_i}\epsilon\dot v_q(t)+\frac{k_d}{k_i}\epsilon^2\ddot v_q(t),\\
&\delta_2(t):=\frac{\epsilon^3}{k_i} v_q^{(3)}(t)-\ddot r(t),\\
&M_0=|\dot r|_\infty+|\ddot r|_\infty, ~~M_1=\big|\dot d\big|_\infty,~~M_2=|g(r,\dot r)|_\infty.
\end{split}
\end{align}

\noindent \textbf{An explanation.} Similar to the proof of Theorem \ref{th1}, the constants $\alpha_i,i=1,\cdots,8$; $c_1$, $c_2$ and $\lambda$ will appear in the following proofs, which depend at most on the constants $L_1,L_2$, $N_1,N_2,\underline b,\bar b,n$ and the triple $(k_p,k_i,k_d)$. We will not explain this property of them one by one later.

Recall $0<\epsilon\le 1$, it is easy to see that there exists some constant $\alpha_0>0$, which is independent of $\epsilon$, such that
\begin{align}\label{46}\sup_{t\geq 0} |\delta_1(t)|\vee |\delta_2(t)|\le  \alpha_0(M_0+M_1).\end{align}
Under notations (\ref{int})-(\ref{pd}), the closed-loop control system (\ref{sys})-(\ref{control}) turns into
\begin{align}\begin{cases}
\mathrm{d}w_0=&\!\frac{1}{\epsilon}w_1\mathrm{d}t\\
\mathrm{d}w_1=&\!\frac{1}{\epsilon}w_2\mathrm{d}t\\
\mathrm{d}w_2=&\!\mathrm{d}x_2+\delta_2\mathrm{d}t.\end{cases}
\end{align}
From (\ref{pe})-(\ref{pd}), it can be deduced that
$$x_1(t)-r(t)= \epsilon w_1(t)-\frac{\epsilon^3}{k_i}\dot v_q(t),~~
 x_2(t)-\dot r(t)=w_2(t)-\frac{\epsilon^3}{k_i}\ddot v_q(t).$$ Besides, recall $f(r(t),\dot r(t),u_r(t))+d(t)=0$,
 it can be obtained that(we omit the time variable $t$ for simplicity in the following derivations)
\begin{align*}
&f(x_1, x_2,u)+d=f(x_1, x_2,u)-f(r,\dot r,u_r)\\
=&A(x_1-r)+B(x_2-\dot r)+\theta(u-u_r)\\
=&\epsilon A w_1+Bw_2-\frac{\theta\bar w}{\epsilon}+\left(\theta\delta_1-\frac{\epsilon^3}{k_i}A\dot v_q-\frac{\epsilon^3}{k_i}B\ddot v_q\right),
\end{align*}
where $A,B$, $\theta$ are $n\times n$ time-varying matrices satisfying
$$\|A\|\le L_1,~~\|B\|\le L_2,~~\underline b\le \lambda_{\min}(\theta)\le \bar b.$$
Similarly, the diffusion term
$$g(x_1,x_2)-g(r,\dot r)=\epsilon C w_1+Dw_2-C\dot v_q\epsilon^3/k_i-D\ddot v_q\epsilon^3/k_i.$$
Consequently, we obtain
\begin{align*}
\mathrm{d}w_2=&\Big(\epsilon A w_1+Bw_2-\frac{\theta \bar w}{\epsilon}+\delta_3\Big)\mathrm{d}t
+(\epsilon C w_1+Dw_2+\delta_4)\mathrm{d}B_t,
\end{align*}
where
\begin{align}\label{53}\begin{split}
&\delta_3:=\theta\delta_1-\frac{\epsilon^3}{k_i}A\dot v_q-\frac{\epsilon^3}{k_i}B\ddot v_q+\delta_2,\\
&\delta_4:= g(r,\dot r)-\frac{\epsilon^3}{k_i}C\dot v_q-\frac{\epsilon^3}{k_i}D\ddot v_q.
\end{split}\end{align}
Denote
$$w=\begin{bmatrix}~w_0~\\~w_1~\\~w_2~\end{bmatrix},~~~\hat A=\begin{bmatrix}
0_n&I_n&0_n\\0_n&0_n&I_n\\
- k_i \theta~&\epsilon^2A- k_p\theta&~\epsilon B- k_d\theta\end{bmatrix},$$
we obtain
\begin{align}
\mathrm{d}w=\frac{\hat Aw}{\epsilon}
 \mathrm{d}t+\begin{bmatrix}0\\0\\ \delta_3\end{bmatrix}\mathrm{d}t+
\begin{bmatrix}0\\0\\ \epsilon C w_1+Dw_2+\delta_4\end{bmatrix}\mathrm{d}B_t.
\end{align}
Similar to the proof of Theorem \ref{th1}, we consider the Lyapunov function $V(w)=w^{\top}Pw$,
where $P$ is a $3n\times 3n$ matrix defined by
 \begin{align}\label{pp}P\overset{\triangle}{=}\frac{1}{2}
\begin{bmatrix}
2\underline bk_ik_pI_n&2\underline bk_ik_dI_n&k_iI_n\\
2\underline bk_ik_dI_n&~(2\underline bk_pk_d-k_i)I_n~&k_p I_n\\
k_iI_n&k_pI_n&k_dI_n
\end{bmatrix}.\end{align}
Then, by some simple manipulations, we have
\begin{align*}
\mathcal{L} V=&\frac{1}{\epsilon}w^\top \left(\hat A^\top P+P\hat A\right)w+\bar w^\top\delta_3+\frac{1}{2}k_d\left|\epsilon C w_1+Dw_2+\delta_4\right|^2\\
=&\underbrace{\frac{1}{\epsilon}w^\top \left(\hat A^\top P+P\hat A\right)w+\frac{1}{2}k_d\left|\epsilon C w_1+Dw_2\right|^2}_{\mathrm{I}}\\
&+\underbrace{\bar w^\top\delta_3+\frac{1}{2}k_d|\delta_4|^2+k_d(\epsilon C w_1+D w_2)^{\top}\delta_4}_{\mathrm{II}},
\end{align*}
where $\bar w=k_iw_0+k_pw_1+k_dw_2.$
From Lemma \ref{lem2},  there exists  $\lambda>0$, which  depends on $k_i,k_p,k_d,L_1,L_2,N_1,N_2,\underline b$ only, such that
$$w^\top \left(\hat A^\top P+P\hat A\right)w+\frac{1}{2}k_d|C\epsilon w_1+Dw_2|^2\le -2\lambda V.$$
Since $0<\epsilon\le 1$, it can be obtained that
\begin{align}\label{I}
\mathrm{I}=&~\frac{1}{\epsilon}w^\top \left(\hat A^\top P+P\hat A\right)w+\frac{1}{2}k_d|C\epsilon w_1+Dw_2|^2\nonumber\\
\le& ~\frac{1}{\epsilon}\Big(w^\top \left(\hat A^\top P+P\hat A\right)w+\frac{1}{2}k_d|C\epsilon w_1+Dw_2|^2\Big)\nonumber\\
 \le& -\frac{2\lambda}{\epsilon} V.
\end{align}
By (\ref{46}), (\ref{53}) and (\ref{vq}), and recall $\theta\le \bar b I_n$,  we conclude that there exists constant $\alpha_1>0$ such that
\begin{align*}
|\delta_3|\le \alpha_1(M_0+M_1),~~
|\delta_4|\le M_2+\alpha_1 (M_0+M_1)\epsilon^3.
\end{align*}
Consequently, there exists $\alpha_2>0$ such that
\begin{align*}
&\left|\bar w^\top\delta_3\right|\le \alpha_2 \sqrt{V}(M_0+M_1), \\
&\ k_d\left|\delta_4\right|^2\le \alpha_2(M_2^2+(M_0+M_1)^2\epsilon^6),\\
&\left|k_d(C\epsilon w_1+D w_2)^{\top}\delta_4\right|\le  \alpha_2\sqrt{V}(M_2+(M_0+M_1)\epsilon^3).
\end{align*}
Therefore, the second term
\begin{align}\label{48}
\mathrm{II}\le & \alpha_3\sqrt{V}\tilde M+\alpha_3\left[M_2^2+(M_0+M_1)^2\epsilon^6\right]\nonumber\\
\le & \frac{\lambda V}{\epsilon}+\frac{\alpha_3^2}{\lambda}\tilde M^2\epsilon+ \alpha_3\left[M_2^2+(M_0+M_1)^2\epsilon^6\right]\nonumber\\
\le & \frac{\lambda V}{\epsilon}+\alpha_4(M_2^2+(M_0+M_1)^2\epsilon),
\end{align}
for some $\alpha_3$ and $\alpha_4$, where $\tilde M:=M_0+M_1+M_2$.
According to (\ref{I})-(\ref{48}), we have the following estimate of $\mathcal{L}V$:
\begin{align*}
\mathcal{L}V\le -\frac{\lambda V}{\epsilon}+ \alpha_4\left[M_2^2+(M_0+M_1)^2\epsilon\right].
\end{align*}
By Lemma \ref{lem4}, we know that
 $$\mathbf{E}V(t)\le V(0)e^{-\lambda t/\epsilon}+\alpha_5\left[ M_2^2\epsilon+(M_0+M_1)^2\epsilon^2\right],$$
which in turn gives \begin{align*}
\mathbf{E}\sum_{i=0}^2|w_i(t)|^2\le&\alpha_6\sum_{i=0}^2|w_i(0)|^2e^{-\lambda t/\epsilon}+\alpha_6\big[M_2^2\epsilon+(M_0+M_1)^2\epsilon^2\big].
\end{align*}
Note that $w_0(0)=\frac{\epsilon v_q(0)}{k_i}$ and \begin{align*}
 w_1(0)\!=\!-\frac{1}{\epsilon}e(0)+\frac{ \epsilon^2 \dot v_q(0)}{k_i},~
w_2(0)\!=\!-\dot e(0)+\frac{ \epsilon^3 \ddot v_q(0)}{k_i},
\end{align*}
and that $0<\epsilon\le 1$, it is not difficult to obtain
\begin{align*}
\mathbf{E}\sum_{i=0}^2|w_i(t)|^2\le& \alpha_7\Big(\frac{|e(0)|^2}{\epsilon^2}+|\dot e(0)|^2+|u_r(0)|^2\epsilon^2\Big)e^{-\lambda t/\epsilon}
\\&+\alpha_7\left(M_2^2\epsilon+(M_0+M_1)^2\epsilon^2\right),
\end{align*}
for some constant $\alpha_7$.
Note that $|e(t)|^2\le 2\epsilon^2 |w_1(t)|^2+\alpha_8\epsilon^6(M_0+M_1)^2,$ and
$|\dot e(t)|^2\le 2|w_2(t)|^2+\alpha_8\epsilon^6(M_0+M_1)^2,$
it can be obtained that
\begin{align*}
\mathbf{E}|e(t)|^2& \le c_1 \Big(|e(0)|^2+|\dot e(0)|^2\epsilon^2+|u_r(0)|^2\epsilon^4\Big)e^{-\lambda t/\epsilon}\\
&+c_2(M_2^2\epsilon^3+(M_0+M_1)^2\epsilon^4),\\
\mathbf{E}|\dot e(t)|^2&\le c_1\Big(\frac{1}{\epsilon^2}|e(0)|^2+|\dot e(0)|^2+|u_r(0)|^2\epsilon^2\Big)e^{-\lambda t/\epsilon}\\
&+c_2(M_2^2\epsilon+(M_0+M_1)^2\epsilon^2).
\end{align*}
Consequently, we have
\begin{align*}
		\limsup_{t\to\infty} \mathbf{E}|e(t)|^2&
\le c_2\left[(M_0+M_1)^2\epsilon+M_2^2\right]\epsilon^3\\
\limsup_{t\to\infty} \mathbf{E}|\dot e(t)|^2&\le  c_2\left[(M_0+M_1)^2\epsilon+M_2^2\right]\epsilon.
	\end{align*}
By definitions of $M_0,M_1$ and $M_2$, it can be seen that the proof of Theorem \ref{th2} is complete.
\hfill $\square$

\textbf{Proof of Theorem \ref{th3}.} We only give the sketch proof, since some details are analogous to proof of Theorem \ref{th2}.

First, we denote $\xi_1(t)=y^*(t)-r(t)$ and $\xi_2(t)=\dot y^*(t)-\dot r(t)$, then it follows from (\ref{filter}) that
\begin{align*}
\dot \xi_1=&\xi_2\\
\dot \xi_2=&-\omega^2 \xi_1-2\omega\xi_2.
\end{align*}
Thus, there exist constants $c_\omega$ and $\lambda_\omega$, which depend on $\omega$ only, such that $|\xi_1(t)|+|\xi_2(t)|\le c_\omega e^{-\lambda_\omega t} (|\xi_1(0)|+|\xi_2(0)|)$. Consequently, we obtain
\[
\sup_{t\geq 0} |\xi_1(t)|+|\xi_2(t)|\le c_\omega (|\xi_1(0)|+|\xi_2(0)|):=\beta_0,
\] where $\beta_0$ depends on the initial states, $r(0)$, $\dot r(0)$ and $\omega$.
Suppose  the PID parameters are determined by formula (\ref{parameter}). Let us denote
\begin{align}\label{EE}\begin{split}
&z_0(t):=-\frac{1}{\epsilon^2}\int_0^t E(s)\mathrm{d}s+\frac{\epsilon }{k_i}v_q(t),\\ &z_1(t):=-\frac{1}{\epsilon}E(t)+\frac{\epsilon^2 }{k_i}\dot v_q(t),\\
&z_2(t):=-\dot E(t)+\frac{\epsilon^3}{k_i}\ddot v_q(t),\\
&\bar z(t):= k_iz_0(t)+k_pz_1(t)+k_dz_2(t),\end{split}
\end{align}
where $E(t)=y^*(t)-x_1(t)$ and $v_q$ is defined in Step 1 in the proof of Theorem \ref{th1}. Then the PID control (\ref{pE}) with parameters given by (\ref{parameter}) can be expressed by
\begin{align}
\hat u=-\frac{\bar z}{\epsilon}+u_r+(v_q-u_r)+\frac{k_p}{k_i}\epsilon\dot v_q+\frac{k_d}{k_i}\epsilon^2\ddot v_q.
\end{align}
Moreover, denote
\begin{align}
\bar \delta_2:=\frac{\epsilon^3}{k_i} v_q^{(3)}-\ddot y^*.
\end{align}
Under  (\ref{EE}), the closed-loop control system turns into
\begin{align}
\mathrm{d}z_0=&\frac{z_1}{\epsilon}\mathrm{d}t\nonumber\\
\mathrm{d}z_1=&\frac{z_2}{\epsilon}\mathrm{d}t\nonumber\\
\mathrm{d}z_2=&\mathrm{d}x_2+\bar \delta_2\mathrm{d}t.
\end{align}
From (\ref{EE}), it can also be deduced that $$x_1= \epsilon y_1-\frac{\epsilon^3}{k_i}\dot v_q+r+\xi_1,~~
 x_2=y_2-\frac{\epsilon^3}{k_i}\ddot v_q+\dot r+\xi_2.$$
Similarly, we obtain
\begin{align*}
\mathrm{d}z_2=&\Big(A\epsilon z_1+Bz_2-\frac{\theta \bar z}{\epsilon}+\bar \delta_3\Big)\mathrm{d}t+(C\epsilon z_1+Dz_2+\bar \delta_4)\mathrm{d}B_t,
\end{align*}
where
\begin{align*}
&\bar \delta_3:=A\xi_1+B\xi_2+\theta\delta_1-\frac{A\epsilon^3\dot v_q}{k_i}-\frac{B\epsilon^3\ddot v_q}{k_i}+\delta_2,\\
&\bar \delta_4:= g(r,\dot r)+C\xi_1+D\xi_2-\frac{C\epsilon^3\dot v_q}{k_i}-\frac{D\epsilon^3\ddot v_q}{k_i},\end{align*}
and $A,B,C,D$ and $\theta$ are $n\times n$ matrices satisfying
$$\|A\|\le L_1,\|B\|\le L_2,\|C\|\le N_1,\|D\|\le N_2,\underline bI_n\le \theta\le \bar bI_n.$$
Denote
$$z=\begin{bmatrix}~z_0~\\~z_1~\\~z_2~\end{bmatrix},~~~\hat A=\begin{bmatrix}
0_n&I_n&0_n\\0_n&0_n&I_n\\
- k_i \theta~&\epsilon^2A- k_p\theta&~\epsilon B- k_d\theta\end{bmatrix},$$ we obtain
\begin{align}
dz=\frac{\hat Az}{\epsilon}
 \mathrm{d}t+\begin{bmatrix}0\\0\\\bar \delta_3\end{bmatrix}\mathrm{d}t+
\begin{bmatrix}0\\0\\ C\epsilon y_1+Dy_2+\bar \delta_4\end{bmatrix}\mathrm{d}B_t.
\end{align}
Similar to the proof of Theorem \ref{th2}, we consider the Lyapunov function $V=z^{\top}Pz$,
where $P$ is given in (\ref{pp}).
Then, by some simple manipulations, we have
\begin{align*}
\mathcal{L} V
=&\frac{1}{\epsilon}z^\top (\hat A^\top P+P\hat A)z+\frac{1}{2}k_d|C\epsilon z_1+Dz_2|^2\\
&+\bar z^\top\bar\delta_3+\frac{1}{2}k_d|\bar\delta_4|^2+k_d(C\epsilon z_1+D z_2)^{\top}\bar \delta_4\\
\le &-\frac{2\lambda}{\epsilon} V+ \beta_1\sqrt{V}+\beta_1\\
\le &-\frac{\lambda}{\epsilon} V+\beta_2
\end{align*}
for some $\beta_1$ and $\beta_2$ independent of $\epsilon.$

By Lemma \ref{lem4}, we know
 $\mathbf{E}V(t)\le V(0)e^{-\lambda t/\epsilon}+\beta_2 \epsilon/\lambda,$
which in turn gives \begin{align*}
\mathbf{E}\sum_{i=0}^2|z_i(t)|^2\le&\beta_3\sum_{i=0}^2|z_i(0)|^2e^{-\lambda t/\epsilon}+\beta_3\epsilon.
\end{align*}
Note that $|z_0(0)|+|z_1(0)|+|z_2(0)|=O(\epsilon)$
and that $0<\epsilon\le 1$, it is not difficult to obtain
\begin{align}\label{zt}
\mathbf{E}\sum_{i=0}^2|z_i(t)|^2\le& \beta_4 \epsilon^2 e^{-\lambda t/\epsilon}
+\beta_3\epsilon.
\end{align}
Note that
$$|E(t)|^2\le 2\epsilon^2 |z_1(t)|^2+O(\epsilon^6),~~\left|\dot E(t)\right|^2\le 2|z_2(t)|^2+O(\epsilon^6),$$
it can be obtained that
\begin{align*}
\mathbf{E}|E(t)|^2& \le c \epsilon^3,~~
\mathbf{E}\left|\dot E(t)\right|^2\le c\epsilon,~~\text{for all}~t\geq 0,
\end{align*}
where the constant $c$ is independent of $\epsilon$(possibly depending on the initial states).

Finally, it can be seen from (\ref{EE}) and (\ref{zt}) that
\begin{align*}
\sup_{t\geq 0}\mathbf{E}|u(t)|^2=&O(\epsilon^{-2})\mathbf{E}\sum_{i=0}^2|z_i(t)|^2+O(1)\\
=&O(\epsilon^{-1}), ~0<\epsilon\le 1.
\end{align*}
This completes the proof of Theorem \ref{th3}.
\hfill $\square$

\section{Conclusions}
In this paper, we have investigated the global tracking problem of the classical PID controller for a basic class of MIMO non-affine nonlinear
uncertain  stochastic systems. In contrast to most of the previous researches, the system model studied in this paper is more general since it considers
unknown dynamics, external disturbances, and time-varying reference signals simultaneously. Explicit design formulae for the PID parameters are
 provided to stabilize the system globally, and the steady-state tracking error is shown to be closely related to properties of reference signals,
   external disturbances, and random noises. Finally, by introducing a desired transient process, we have also presented a new PID tuning rule,
    which can guarantee both nice steady-state and superior transient control performances. Of course, many important issues are still open and
     worth investigating. For example, it would be interesting and meaningful to consider the situation with measurement noises and input saturation,
      and to generalize the results in this paper to more general dynamical systems.

\section{Appendix}
Proof of Lemma \ref{lem1}. We first prove the case $n=1$.
Let $J$ be a nonnegative function belonging to $C_0^\infty(\mathbb{R})$ and have the following properties:

(i) $~\!J(t)=0$~ if $~|t|\geq 1$,

(ii) $~\!\int_{-\infty}^\infty J(t)\mathrm{d}t=1$.

For example, we may take
\begin{align}
J(t)=\begin{cases}ke^{-\frac{1}{1-t^2}}, ~~~~\text{if}~ |t|<1,\\
0,~~~~~~~~~~  ~~~\!~\!\text{if}~ |t|\geq1,
\end{cases}
\end{align}
where $k>0$ is chosen such that condition (ii) is satisfied.
Now, for $h\in C^1_b([0,\infty),\mathbb{R}^1)$, we first extend $h$ to a function defined on $\mathbb{R}$ by
$h(t)=2h(0)-h(-t)$, for $t<0$. Then, we define $p(t)$ as follows:
\begin{align}\label{pt}
p(t):=\int_{-\infty}^\infty h(s)J(t-s)\mathrm{d}s=\int_{-1}^1 h(t-s)J(s)\mathrm{d}s.
\end{align}
By the property of convolution, it is well-known that $p(t)\in C^{\infty}([0,\infty),\mathbb{R}^1)$. Next, it is easy to see
\begin{align*}
|h(t)-p(t)|=&\left|\int_{-1}^1 (h(t)-h(t-s))J(s)\mathrm{d}s\right|\\
\le&\int_{-1}^1 |h(t)-h(t-s)|J(s)\mathrm{d}s\\
\le& \sup_{t\geq 0}\left|\dot h(t)\right|\int_{-1}^1 J(s)\mathrm{d}s\\
=&\sup_{t\geq 0}\left|\dot h(t)\right|,~~\text{for all}~t\geq 0.
\end{align*}
Besides, from (\ref{pt}), it can be deduced that
\begin{align*}
|\dot p(t)|=&\left|\int_{-1}^1 \dot h(t-s)J(s)\mathrm{d}s\right|\\
\le &\sup_{t\geq 0}\left|\dot h(t)\right| \int_{-1}^1 J(s)\mathrm{d}s\\=&\sup_{t\geq 0}\left|\dot h(t)\right|,~\text{for all}~t\geq 0.
\end{align*}
We next estimate the upper bounds of $|\ddot p(t)|$ and $\big|p^{(3)}(t)\big|$.
From the identities $$\dot p(t)=\int_{-\infty}^\infty\dot h(t-s)J(s)\mathrm{d}s=\int_{-\infty}^\infty \dot h(s)J(t-s)\mathrm{d}s,$$
we conclude that
\begin{align*}
\ddot p(t)=&\int_{-\infty}^\infty \dot h(s)\dot J(t-s)\mathrm{d}s,\\
p^{(3)}(t)=&\int_{-\infty}^\infty \dot h(s)\ddot J(t-s)\mathrm{d}s.
\end{align*}
Consequently, we have
\begin{align*}
&|\ddot p(t)|\le \left[\int_{-\infty}^\infty \left|\dot J(s)\right|\mathrm{d}s \right]\sup_{t\geq 0}\left|\dot h(t)\right| ,\\
&\left|p^{(3)}(t)\right|\le  \left[\int_{-\infty}^\infty \left|\ddot J(s)\right|\mathrm{d}s\right] \sup_{t\geq 0}\left|\dot h(t)\right|.
\end{align*}
Let us choose $$c_1\overset{\triangle}{=}\max\left\{1, ~\int_{-\infty}^\infty \left|\dot J(t)\right|\mathrm{d}t,~~ \int_{-\infty}^\infty \left|\ddot J(t)\right|\mathrm{d}t \right\},$$ then (\ref{distance}) is satisfied for $n=1.$

Next, we consider the case $n\geq 1$. Suppose $h=[h_1,\cdots,h_n]\in C_b^1([0,\infty),\mathbb{R}^n)$. For each $i$, we know that there exists $p_i\in C_b^3([0,\infty),\mathbb{R}^1)$, which satisfies
\begin{align}\label{11}
\begin{split}
&|h_i(t)-p_i(t)|\vee \left|\dot p_i(t)\right|\vee \left |\ddot p_i(t)\right|\vee\left|p_i^{(3)}(t)\right| \\
\le &~c_1\sup_{t\geq 0}\left|\dot h_i(t)\right|,~~\text{for all}~~t\geq 0,~i=1,\cdots,n.\end{split}
\end{align}
Now, set $p(t)=[p_1(t),\cdots,p_n(t)]$, then $p\in C_b^3([0,\infty),\mathbb{R}^n)$. Moreover, it follows from (\ref{11}) that
\begin{align*}
|h(t)-p(t)|=&\left(\sum_{i=1}^n \left|h_i(t)-p_i(t)\right|^2\right)^{\frac{1}{2}}
\le \left(\sum_{i=1}^n c_1^2\sup_{t\geq 0}\left|\dot{h}_i(t)\right|^2  \right)^{\frac{1}{2}}\\
\le &\left(\sum_{i=1}^n c_1^2\sup_{t\geq 0}\left|\dot{h}(t)\right|^2  \right)^{\frac{1}{2}}
=c_1\sqrt{n}\sup_{t\geq 0}\left|\dot{h}(t)\right|.
\end{align*}
Similarly, from (\ref{11}), it can be obtained that
\begin{align*}
\begin{split}
 \left|\dot p(t)\right|\vee \left |\ddot p(t)\right|\vee\left|p^{(3)}(t)\right|
\le c_n\sup_{t\geq 0}\left|\dot h(t)\right|,~\text{for all}~t\geq 0,\end{split}
\end{align*}
where the constant $c_n$ can be taken as $c_1\sqrt{n}$. \hfill$\square$

Proof of Lemma \ref{lem2}.

\emph{Step 1.} We first prove that $P>0$. Denote $B=\left[2\underline b K_iK_d~~K_i\right]$ and
\begin{align*}D=2\underline b K_iK_p, ~~E=
	\begin{bmatrix}
		2\underline bK_pK_d-K_i~&K_p\\
		K_p&K_d
	\end{bmatrix},
\end{align*}
then $P=\frac{1}{2}\begin{bmatrix}
	D&B\\
	B^{\top}&E
\end{bmatrix}$.
Note that $D>0$, to prove $P>0$,  by applying  \cite[Theorem 7.7.6]{horn}, it suffices to show that
$E-B^{\mathsf{T}}D^{-1}B>0$. Note that $K_p,$ $K_i$ and $K_d$ are diagonal matrices(and thus they are commutative), it can be calculated that
\begin{align*}&E-B^{\mathsf{T}}D^{-1}B\\
	=&
	\begin{bmatrix}
		2\underline b(K_p^2-K_iK_d)K_p^{-1}K_d-K_i&(K_p^2-K_iK_d)K_p^{-1}\\
		~~(K_p^2-K_iK_d)K_p^{-1}&K_d-K_iK_p^{-1}/(2\underline b)
	\end{bmatrix}\\
	\overset{\triangle}{=}&
	\begin{bmatrix}
		D_1&B_1\\
		B_1&E_1
	\end{bmatrix}.
\end{align*}
To prove $E-B^{\mathsf{T}}D^{-1}B>0$, by applying Theorem 7.7.6 in \cite{horn} again, we only need to show $D_1>0,$ and
\begin{align}\label{32}
	E_1-B_1^{\mathsf{T}}D_1^{-1}B_1>0.
\end{align}
Firstly, from (\ref{th}), we know that $K_p^2>2K_iK_d$ and $K_d^2>K_p/\underline b$, hence
\begin{align*}
	D_1=&2\underline b(K_p^2-K_iK_d)K_p^{-1}K_d-K_i\\
	>&2\underline bK_iK_d^2K_p^{-1}-K_i=(2\underline bK_d^2-K_p)K_iK_p^{-1}\\
	>&\underline bK_d^2K_iK_p^{-1}>0.
\end{align*}
Secondly, note that $D_1$, $B_1$ and $E_1$ are all $n\times n$ diagonal matrices,  thus (\ref{32}) is equivalent to
$E_1D_1-B_1^2>0.$
By some standard  calculations,
\begin{align*}
	&E_1D_1-B_1^2
	=K_p^{-1}\left(4K_p^2K_d^2\underline b^2+K_i^2-2K_p^3\underline b-4K_iK_d^3\underline b^2\right)/(2\underline b).
\end{align*}
Write \begin{align*}
K_p=\text{diag}(k_{p1},k_{p2},\cdots,k_{pd}),~	K_i=\text{diag}(k_{i1},\cdots,k_{id}),~
K_d=\text{diag}(k_{d1},k_{d2},\cdots,k_{dd}),
\end{align*}
from $K_p^2>2K_iK_d$ and $K_d^2>K_p/\underline b$, we know that
$k_{pj}^2>2k_{ij}k_{dj}$ and $k_{dj}^2>k_{pj}/\underline b~$ for $j=1,2,\cdots,n$. Therefore, based on the proof of \cite[Proposition 4.2]{2022},  we conclude that the $j$-th diagonal element of the matrix $4K_p^2K_d^2\underline b^2+K_i^2-2K_p^3\underline b-4K_iK_d^3\underline b^2$ is positive. Hence, $E_1D_1-B_1^2>0$.

\emph{Step 2.} By some simple calculations, it can be deduced that
\begin{align*}
\mathrm{I}:=&2Y^{\top}P\begin{bmatrix}y_1\\y_2\\Ay_1+By_2-\theta\bar y\end{bmatrix}\\
=&2\underline b y_0^{\top}K_iK_py_1+2\underline b y_0^{\top}K_iK_dy_2+2\underline by_1^{\top}K_iK_dy_1\nonumber\\
&+y_1^{\top}K_iy_2+y_1^{\top}(2\underline bK_pK_d-K_i)y_2+y_2^{\top}K_py_2\\
&+\bar y^{\top}(Ay_1+By_2-\theta\bar y).
\end{align*}
First, note that $\theta \geq \underline b I_n$ and $\bar y=K_iy_0+K_py_1+K_d y_2$, we can obtain
\begin{align}
&-\bar y^{\top}\theta\bar y\le-\underline b \left|K_iy_0+K_py_1+K_dy_2\right|^2\nonumber\\
=&-\underline b\big(y_0^{\top}K_i^2y_0+y_1^{\top}K_p^2y_1+y_2^{\top}K_d^2y_2\nonumber\\
&+2y_0^{\top}K_iK_py_1+2y_0^{\top}K_iK_dy_2+2y_1^{\top}K_pK_dy_2\big).
\end{align}
Consequently, we have
\begin{align}\begin{split}
\mathrm{I}\le &-\underline b y_0^{\top}K_i^2y_0-\underline b y_1^{\top}(K_p^2-2K_iK_d)y_1\\
&-y_2^{\top}(\underline b K_d^2-K_p)y_2+\bar y^{\top}(Ay_1+By_2).\end{split}
\end{align}
Next, notice $\|A\|\le L_1,~\|B\|\le L_2$, it can be obtained that
\begin{align*}
&y_0^{\top}K_i(A y_1+By_2)\le \frac{\underline b \|K_i\|^2}{2}|y_0|^2+\frac{L_1^2}{\underline b}|y_1|^2+\frac{L_2^2}{\underline b}|y_2|^2,\end{align*}
and \begin{align*}
&\left(y_1^{\top}\!K_p\!+y_2^{\top}\!K_d\right)(A y_1+By_2)\\
\le & \big(\|K_p\|L_1+\hat k/2\big)|y_1|^2+\big(\|K_d\|L_2+\hat k/2\big)|y_2|^2,
\end{align*}
where $\hat k:=\|K_p\| L_2+\|K_d\| L_1$.
From (\ref{th}), it is not difficult to obtain
\begin{align}\label{bar k}\begin{split}
&\|K_p\|^2>\lambda_{\text{m}}(K_p^2-2K_iK_d)>\bar k,\\
&\|K_d\|^2>\lambda_{\text{m}}(K_d^2-K_p/\underline b)>\bar k,\end{split}
\end{align}
where $\bar k=(\|K_p\|+\|K_d\|)(L_1+L_2)/\underline b$. From (\ref{bar k}), one can obtain the following inequalities:
\begin{align}\label{kpkd}\|K_p\|\geq 2(L_1+L_2)/\underline b,~~\|K_d\|\geq 2(L_1+L_2)/\underline b.\end{align}
By the definitions of $\hat k$ and $\bar k$, it follows from  (\ref{kpkd}) that
\begin{align*}
\|K_p\|L_1+\hat k/2+L_1^2/\underline b\le \bar k,~
\|K_d\|L_2+\hat k/2+L_2^2/\underline b\le \bar k.
\end{align*}
Consequently, we obtain
\begin{align}\begin{split}
\bar y^{\top}(A y_1+By_2)\le \frac{\underline b\|K_i\|^2|y_0|^2}{2}
+\bar k \underline b (|y_1|^2+|y_2|^2).\end{split}
\end{align}
Therefore, the first term satisfies
\begin{align} \label{i}
\text{I}\le& -\underline b\Big(y_0^{\top}\left(K_i^2-\|K_i\|^2I_n/2\right)y_0+y_1^{\top}(K_p^2-2K_iK_d)y_1\nonumber\\
&+y_2^{\top}(K_d^2-K_p/\underline b)y_2\Big)
+\bar k\underline b(|y_1|^2\!+\!|y_2|^2).
\end{align}
On the other hand, since $\|C\|\le N_1,~\|D\|\le N_2$, thus the second term  has the following upper bound:
\begin{align}\label{iii}\begin{split}
\mathrm{II}:=\frac{1}{2}\|K_d\||C y_1+D y_2|^2
\le \|K_d\|\sum_{i=1}^2N_i^2|y_i|^2.\end{split}
\end{align}
Combine (\ref{i}) and (\ref{iii}), we conclude that
\begin{align*}
\mathrm{I}+\mathrm{II}\le  & -\underline b y_0^{\top}\Big(K_i^2-\frac{\|K_i\|^2I_n}{2}\Big)y_0\nonumber\\
&-\underline b y_1^{\top}\left[K_p^2-2K_iK_d-\left(\bar k+\|K_d \|N_1^2/\underline b\right)I_n\right]y_1 \nonumber\\
&-\underline by_2^{\top}\left[K_d^2-K_p/\underline b-\left(\bar k+\|K_d \|N_2^2/\underline b\right)I_n\right]y_2.
\end{align*}
By inequalities (\ref{th}), there exists $\gamma>0$ which depends on $(K_i,K_p,K_d,L_i,N_i,\underline b)$ only, such that
\begin{align*}
\mathrm{I}+\mathrm{II}\le  &-\gamma |Y|^2, ~\text{for all}~ Y=[y_0^\top,y_1^\top,y_2^\top]^\top\in\mathbb{R}^{3n}.
\end{align*}
Therefore, the proof of Lemma \ref{lem2} is complete. \hfill$\square$

\bibliographystyle{plain}

\end{document}